\documentclass{PoS}

\title{Charged Hadron Properties\\ in Background Electric Fields}

\ShortTitle{Hadrons in Background $\mathcal{E}$-Fields}

\author{Will Detmold\\
        Department of Physics\\
        College of William and Mary\\
        Williamsburg, VA, 23187, USA\\
        Jefferson Laboratory\\
        12000 Jefferson Avenue\\
        Newport News, VA 23606, USA\\
        E-mail: \email{wdetmold@wm.edu}
}
\author{Brian Tiburzi\\
        Department of Physics\\
        University of Maryland\\
        College Park, MD 20742, USA\\
        E-mail: \email{bctiburz@umd.edu}
}
\author{\speaker{Andr\'{e} Walker-Loud}\\
        Department of Physics\\
        College of William and Mary\\
        Williamsburg, VA, 23187, USA\\
        E-mail: \email{walkloud@wm.edu}}

\abstract{We report on a lattice calculation~\cite{Detmold:2009dx} demonstrating a novel new method to extract the electric polarizability of charged pseudo-scalar mesons by analyzing two point correlation functions computed in classical background electric fields.}

\FullConference{The XXVII International Symposium on Lattice Field Theory - LAT2009\\
		 July 26-31 2009\\
		 Peking University, Beijing, China}

\begin{document}

\def\eqref#1{{(\ref{#1})}}

\section{Introduction}
A staple component of any electrodynamics or quantum mechanics course is the electric polarizability.  Neutral material immersed in a weak external field polarizes, internally setting up an electric dipole moment, aligned so as to minimize the energy.  At the atomic level, the electron clouds are distorted creating these microscopic dipole moments.  The same process occurs at the hadronic level but the polarization effects are now constrained by the strong force.  Polarizabilities of these bound QCD states can be viewed as a distortion of the charged pion cloud of a given hadron.

One can use lattice QCD to non-perturbatively compute the quark and gluon interactions in the presence of background electric (or magnetic) fields.  For sufficiently weak background fields, the low energy properties of the hadrons can be rigorously computed using effective field theory.  With this treatment, a picture of hadrons emerges from chiral dynamics: 
that of a hadronic core surrounded by a pseudoscalar meson cloud.  As some pseudoscalar mesons are charged, polarizabilities of hadrons encode the stiffness of the charged meson cloud (as well as that of the core).  The form of pseudoscalar meson polarizabilities is consequently strongly constrained by chiral dynamics%
~\cite{Holstein:1990qy,Burgi:1996qi,Gasser:2006qa}.
However, beyond the leading order, the results depend upon essentially unknown low-energy constants, which must currently be estimated in a \textit{model-dependent} fashion.  In the case of the charged pion, the experimental measurement of the polarizability has proven difficult, both in the original measurement~\cite{Antipov:1982kz} as well as the most recent published result~\cite{Ahrens:2004mg}.  Currently, there is a 2-3 sigma discrepancy between the two-loop $\chi$PT prediction and the measured charged pion polarizability.  New results with higher statistics and the first kaon results are anticipated from COMPASS at CERN~\cite{Abbon:2007pq}.

A lattice QCD calculation of these quantities is thus very timely.  This will allow us to both:
\begin{itemize}
\item make predictions/compare with new experimental results
\item explore stringent predictions from chiral dynamics.  For example, the pion polarizabilites (from $\chi$PT)
\begin{eqnarray}\label{eq:ChPTPols}
	\alpha_E^{\pi^\pm} &=& \frac{8 \alpha_{f.s.}}{m_\pi f_\pi^2} \left( L_9 + L_{10} \right)\, , \nonumber\\
	\alpha_E^{\pi^0} &=& -\frac{\alpha_{f.s.}}{3 m_\pi (4\pi f_\pi)^2}\, .
\end{eqnarray}
\end{itemize}

\section{Background Field Implementation}
To produce a constant electric field, $\vec{\mathcal{E}} = \mathcal{E} \hat{z}$, we use the Euclidean space vector potential,
\begin{equation}
	A_\mu(x) = (0,0,-\mathcal{E} t, 0)\, ,
\end{equation}
where $\mathcal{E}$ is a real-valued parameter.  The analytic continuation $\mathcal{E} \rightarrow -i \mathcal{E}_M$ produces a real-valued electric field in Minkowski space.  This analytic continuation can not generally be performed with numerical data due to non-perturbative effects, such as Schwinger pair-creation, which is absent in Euclidean space.  However, we are only interested in quantities which are perturbative in the field-strength, for which the na\"{i}ve continuation produces the correct Minkowski-space phsyics, see Ref.~\cite{Tiburzi:2008ma} for explicit details.

\begin{figure}
\begin{tabular}{cc}
\includegraphics[width=0.45\textwidth]{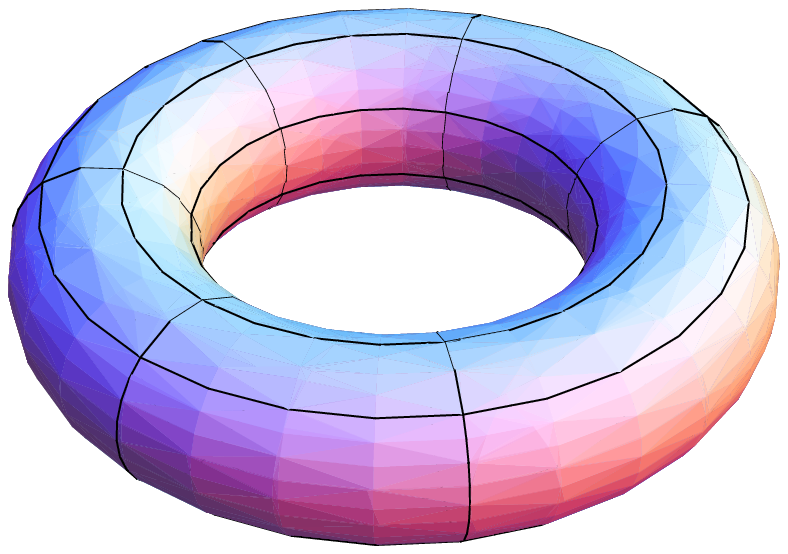}
&
\includegraphics[width=0.35\textwidth]{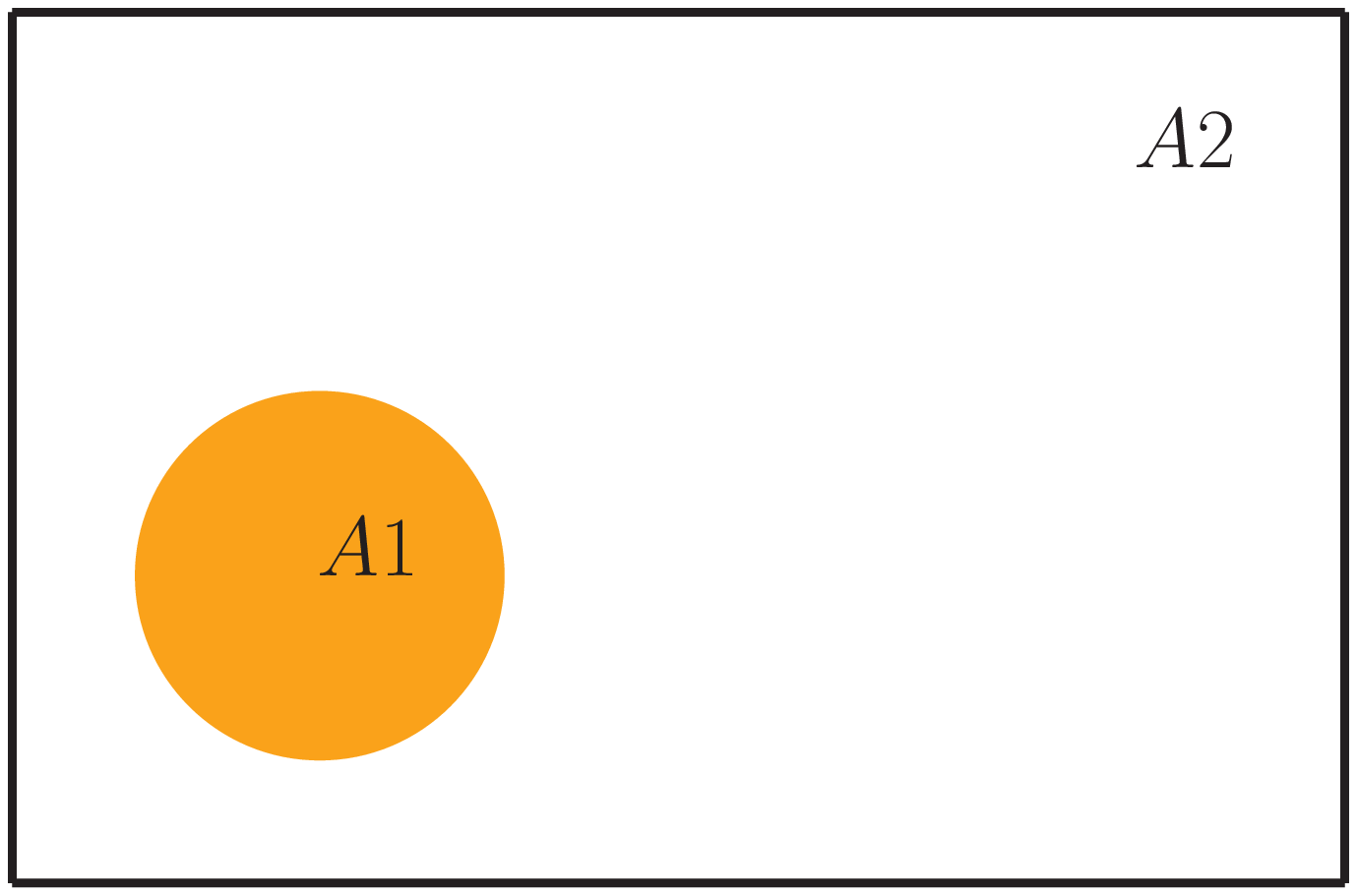}
\\
&$\qquad \Phi_1 = -\Phi_2 + 2\pi n,\quad n\in Z$\\
$(a)$&$(b)$
\end{tabular}
\caption{\label{fig:torus}For a background gauge-potential implemented on a periodic torus $(a)$, the gauge-flux through area $A1$ and $A2$ $(b)$ must be equal and opposite in order to have well defined, single valued action.  This directly leads to the quantization condition, Eq.~(2.3).}
\end{figure}

For our initial studies, we implement a quenched electric field (the valence quarks are given an electric charge while the sea quarks remain neutral) by post-multiplying existing $SU(3)$ gauge links with the Abelian links,
\begin{equation}
U_\mu(x) \rightarrow U_\mu(x) U_\mu^{(\mathcal{E})}(x),
\qquad\qquad
U_\mu^{(\mathcal{E})}(x) = \exp \left\{ iQ A_\mu(x) \right\}.
\end{equation}
On a torus, construction of a constant field requires quantization to maintain a single-valued action~\cite{'tHooft:1979uj,'tHooft:1981sz,vanBaal:1982ag}.  
The argument is as follows.  An action with periodic boundary conditions is defined on a torus, which is a closed surface, Fig.~\ref{fig:torus}$(a)$.  Therefore, to maintain a single valued action, the net flux through the $x_3$--$x_4$ plane, Fig.~\ref{fig:torus}$(b)$ must be zero, modulo $2\pi$, i.e. $\Phi = Q \mathcal{E} \beta L \equiv 2\pi n$, with $n \in Z$, immediately yielding the 't Hooft quantization condition
\begin{equation}\label{eq:Quant}
	\mathcal{E} = 
	\frac{ 2 \pi n} { q_d \, \beta L}\, .
\end{equation}
Use of the down quark charge $q_d = -1/3 e$, will lead to the up quark, with charge $q_u = 2/3 e$ automatically satisfying Eq.~\eqref{eq:Quant}.

The implementation of a constant gauge field must be slightly modified for a discrete torus~\cite{Smit:1986fn,Rubinstein:1995hc,AlHashimi:2008hr}.
  See Refs.~\cite{Detmold:2009dx,Detmold:2008xk,Detmold:2009fr} for our explicit implementation as well as a quantitative study the effects on the pion spectra from using non-quantized field values.

\section{Lattice Calculation}
For our calculation, we have used the $2+1$ flavor $20^3\times 128$ dynamical anisotropic gauge ensembles with a renormalized anisotropy of $\xi = a_s / a_t = 3.5$ and a pion mass of $m_\pi \sim 390$~\texttt{MeV}~\cite{Lin:2008pr} generated by the Spectrum Collaboration.  We used 200 configurations, spaced by either 20 or 40 monte-carlo time units.  On each configuration, for each background field strength, we computed an average of 10 \textit{up}, \textit{down} and \textit{strange} propagators with random spatial sources using the recently developed \texttt{eigCG} deflation algorithm~\cite{Stathopoulos:2007zi}.  For this work, we used the background field strengths given by Eq.~\eqref{eq:Quant}, for $n = 0, \pm1, \pm2, \pm3, \pm4$.  On our lattices, this yields the expansion parameter governing the deformation of a hadron's pion cloud~\cite{Tiburzi:2008ma}
\begin{equation}
	\left( \frac{e \mathcal{E}}{m_\pi^2} \right) = 0.18 n^2\, ,
\end{equation}
leading us to expect the need of including terms beyond quadratic order in $\mathcal{E}$ to describe the modified hadron energies.

\section{Charged Particle Correlation Functions in Background $\mathcal{E}$-fields}
For neutral particles, it is straightforward to calculate the electric polarizability using standard spectroscopy~\cite{Fiebig:1988en}.  The lattice two-point functions have their standard forms with the overlap factors and energies now depending upon the external field strength, $\mathcal{E}$.  For charged particles, the correlation functions are more complicated.  Consider a relativistic scalar particle of charge $Q$.  As the particle is composite, there are both Born and non-Born terms in the action.  The non-Born terms account for non-minimal couplings of the field to the particle, such as the polarizabilities.  These couplings can be summed, as in the case of neutral particles, into the \textit{energy} of the particle, $E(\mathcal{E})$.%
\footnote{The quantity $E(\mathcal{E})$ no longer has the interpretation of the energy of the hadron, but remains given by Eq.~\eqref{eq:MofE}.} 
Additionally, to describe the two-point correlation functions, one must also sum the Born couplings, arriving at~\cite{Tiburzi:2008ma}
\begin{equation} \label{eq:GCharged}
G(t, \mathcal{E}) =
	Z(\mathcal{E}) \, D\big(t, E(\mathcal{E}), \mathcal{E} \big) +Z'(\mathcal{E}) \, D\big(t, E'(\mathcal{E}), \mathcal{E} \big)
	+ \ldots ,
\end{equation} 
where
\begin{equation} \label{eq:QProp}
D\big(t, E(\mathcal{E}), \mathcal{E}\big) = \int_0^\infty ds \sqrt{ \frac{ Q \mathcal{E} }{2 \pi \sinh ( Q \mathcal{E}  s ) } }
	\exp \left[
		- \frac{ Q \mathcal{E}  t^2}{2} \coth (  Q \mathcal{E}   s )
		- \frac{E(\mathcal{E})^2 s}{2}
	\right]\, ,
\end{equation}
and
\begin{equation}\label{eq:MofE}
E(\mathcal{E}) = M 
	+ \frac{1}{2}4\pi \alpha_E \mathcal{E}^2
	-\frac{1}{4!}(4\pi)^2 \bar{\alpha}_{EEE} \mathcal{E}^4
	+\dots
\end{equation}
For sufficiently weak fields, Eq.~\eqref{eq:QProp} reduces to that in Ref.~\cite{Detmold:2006vu}
\begin{equation}
D\big(t, E(\mathcal{E}), \mathcal{E}\big) \rightarrow
	\frac{1}{2E(\mathcal{E})} 
	\exp \left\{ -E(\mathcal{E}) t 
		- \frac{\mathcal{E}^2}{E(\mathcal{E})^4} \left(
			\frac{1}{6} (E(\mathcal{E}) t)^3 + \frac{1}{4}(E(\mathcal{E}) t)^2
			+\frac{1}{4}(E(\mathcal{E}) t) \right)
	\right\}
	+\cdots
\end{equation}

\section{Results}
\begin{figure}
\center
\includegraphics[width=0.45\textwidth]{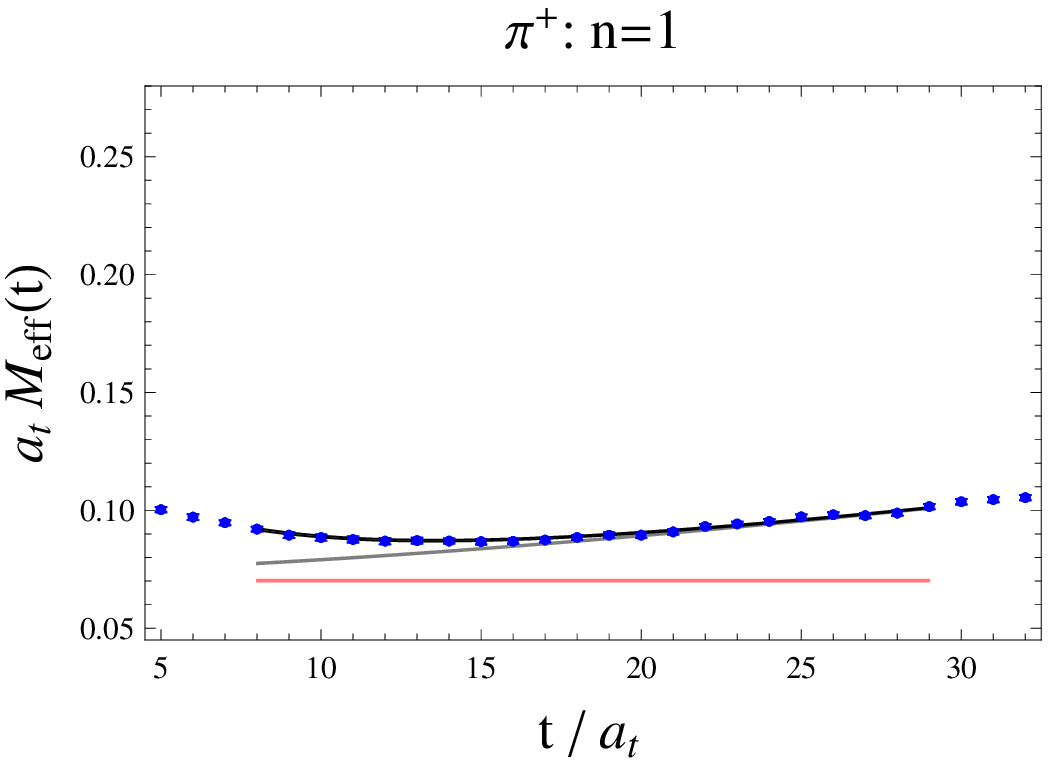}
\includegraphics[width=0.45\textwidth]{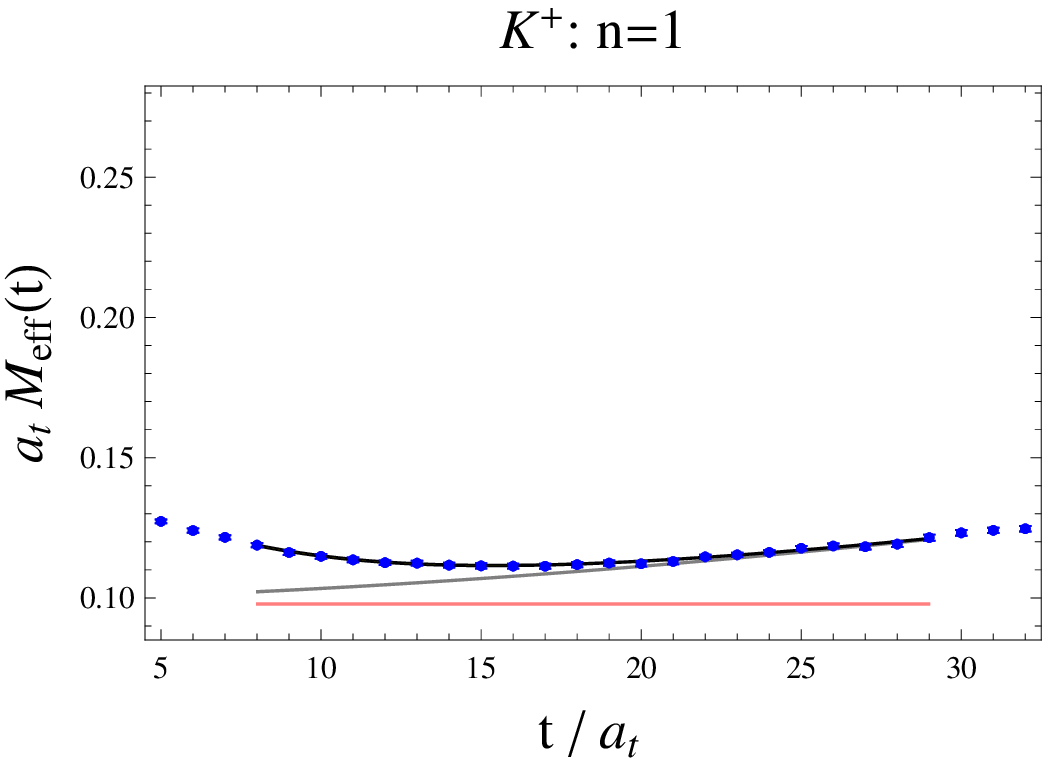}
\includegraphics[width=0.45\textwidth]{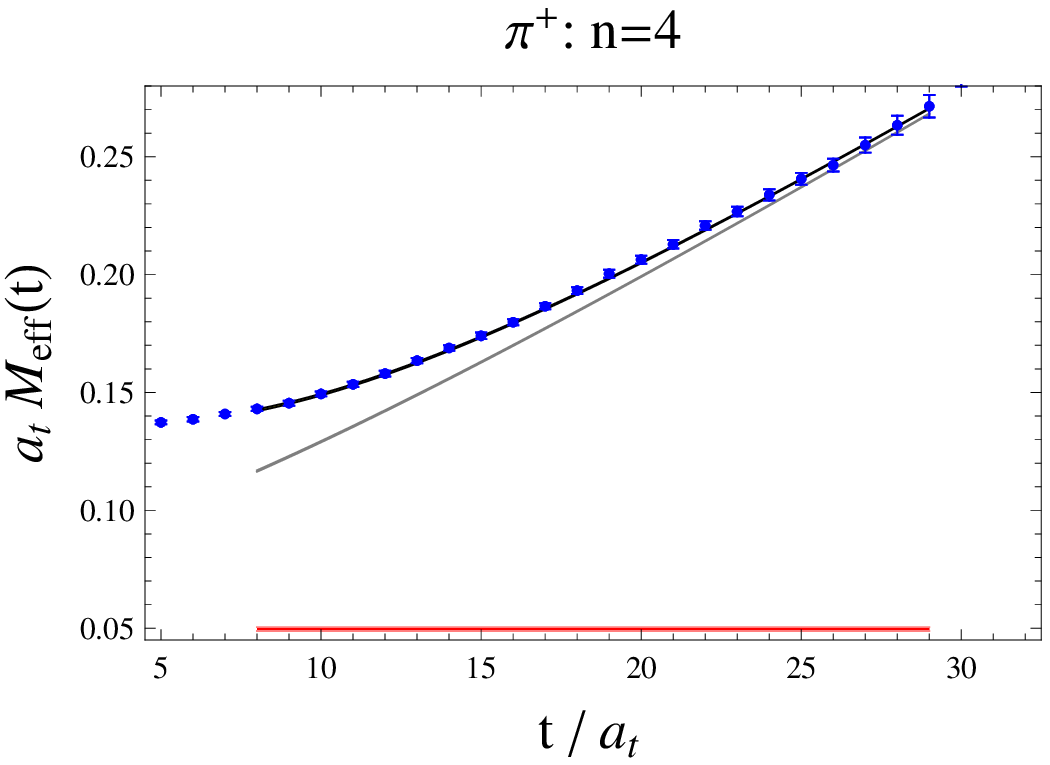}
\includegraphics[width=0.45\textwidth]{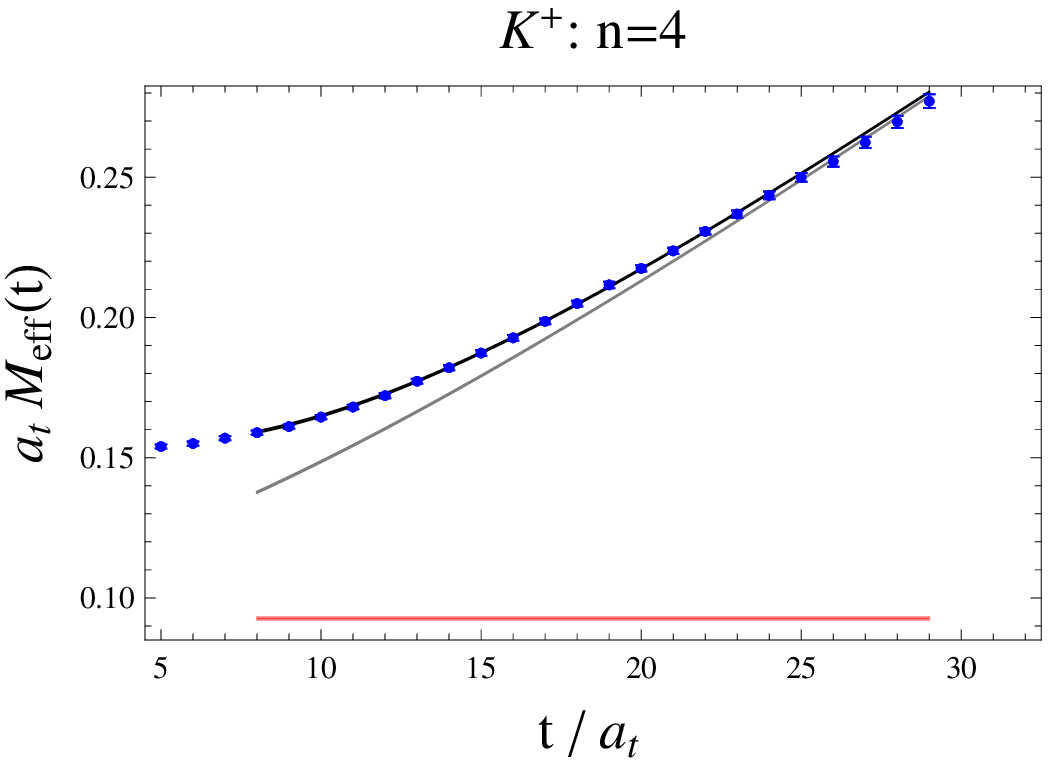}
\caption{\label{fig:effMass} Example effective mass plots of charged pion correlation functions, $a_t M_{eff} = \ln ( G(t) / G(t+1) )$.}
\end{figure}

In this section, we first present the results of our charged pseudo-scalar mass determinations from the long-time behavior of the two-point correlation functions calculated in the presence of background electric fields.  We then present the resulting determination of the $\pi^\pm$ and $K^\pm$ electric polarizabilites, determined from a fit of our extracted masses to Eq.~\eqref{eq:MofE}.  The anisotropy of the ensembles used provided a sufficiently fine time resolution as to allow for a fit to the correlation functions using a two-state fit, which in turn reduced the systematic uncertainty arising from the choice of fitting window.
In Fig.~\ref{fig:effMass}, we display sample fits to the charged pion and kaon correlation functions.  The blue data points are the resulting effective masses,
\begin{equation}
	 a_t M_\textrm{eff}(t) = \ln \left( \frac{G(t)}{G(t+1)} \right)\, .
\end{equation}
The yellow band (spanning $t=8-29$) is the 68\% confidence interval using the two-state fit with Eqs.~\eqref{eq:GCharged} and \eqref{eq:QProp}.  The gray band is the reconstructed (effective mass of the) correlation function using only the ground state correlator and the horizontal salmon colored band is the resulting ground state energy, $E(\mathcal{E})$ which enters Eq.~\eqref{eq:QProp}.

\begin{figure}[b]
\center
\includegraphics[width=0.45\textwidth]{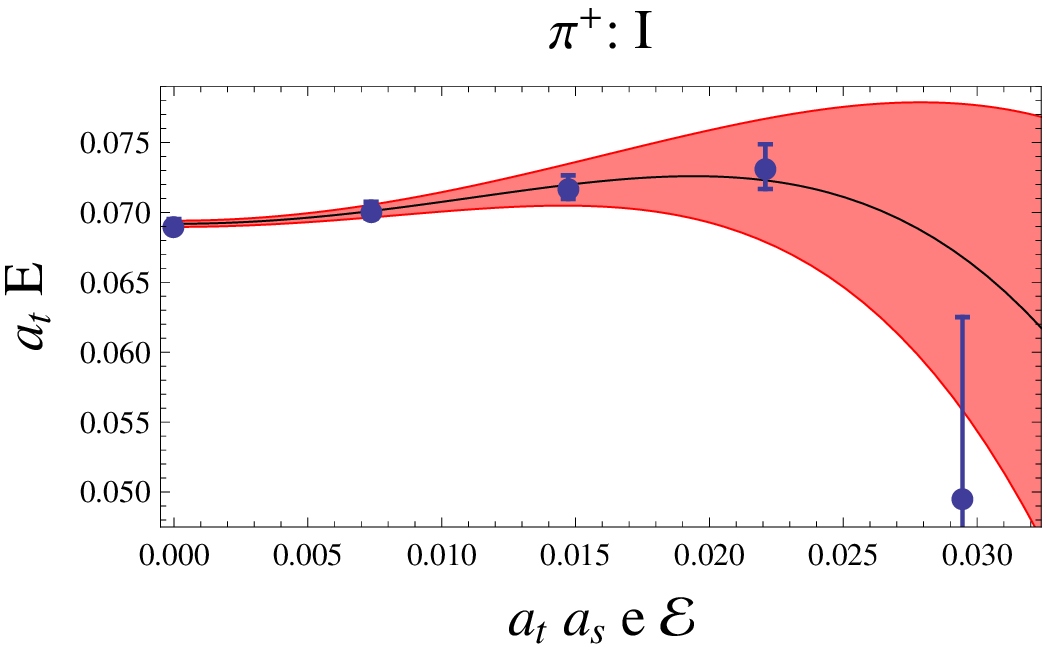}
\includegraphics[width=0.45\textwidth]{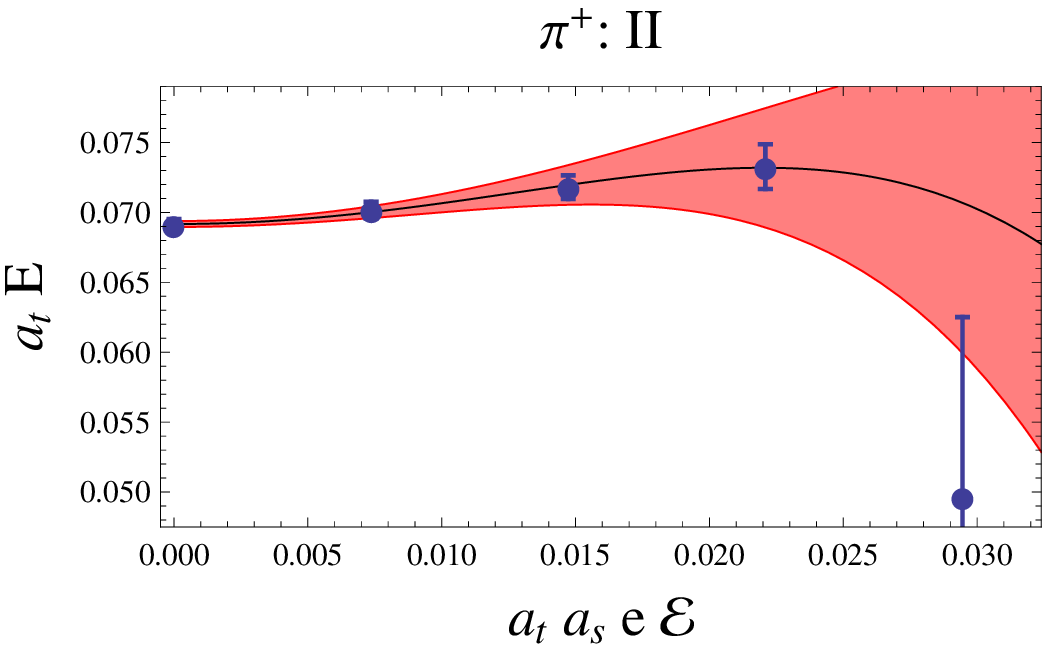}
\caption{\label{fig:pols}The fits to the energy are to all values of $n$ (I) and up to $n= \pm 3$ (II).  $1-P$ is the quality of fit.}
\end{figure}

In the upper section of Table~\ref{tab:masses}, we collect the results of the ground state energies determined from this analysis.  To enforce the invariance under parity transformations, for each $n$ (from $\mathcal{E} = 2\pi n / q_d \beta L$) we have taken the average energy from $n = \pm 1, \pm2$ etc.
In the bottom half of Table~\ref{tab:masses}, we present the resulting polarizabilites extracted from fitting the pseudo-scalar energies to Eq.~\eqref{eq:MofE}.  In Fig.~\ref{fig:pols}, we present the resulting energy dependence as a function of the background field, and the resulting 68\% confidence band.  The first fit (I) utilizes all five valued of $\mathcal{E}$ while the second fit (II) results from a fit to the smallest four background field strengths.

\begin{table}
\caption{\label{tab:masses}Summary of fit results for charged meson two-point functions for $8 \leq t \leq  29$.}
\center
\begin{tabular}{cccc||cccc}
\hline\hline
$$ & $n$ & $a_t E(\mathcal{E})$ & $1-P$ & $$ & $n$ & $a_t E(\mathcal{E})$ & $1-P$ 
\tabularnewline
\hline
$\quad \pi^+ \quad$ & $0$ & $0.0691(4) $ & $0.66$ &
$\quad K^+ \quad$  & $0$ & $0.0969(3) $ & $0.70 $ 
\tabularnewline
& $1$ & $0.0702(6) $ &$0.46$ &
& $1$ & $0.0979(4)$ & $0.77$
\tabularnewline
& $2$ & $0.0718(8) $ & $0.61$ &
& $2$ & $0.0982(7)$ & $0.78$
\tabularnewline
& $3$ & $0.0733(16) $ & $0.93$ &
& $3$ & $0.0958(10) $  & $0.98$
\tabularnewline
& $4$ & $0.0497(129) $ & $0.97$ &
& $4$ & $0.0927(23) $ & $0.97$ 
\tabularnewline
\hline
\hline
\tabularnewline
\end{tabular}
\begin{tabular}{ccccc}
\hline\hline
$\pi^+ $ & $\quad  a_t M \quad $ &  $\quad \alpha_E^{\textrm{latt}} \quad  $  & $\quad \overline{\alpha}^{\textrm{latt}}_{EEE} \quad $ & $1-P $ 
\tabularnewline
\hline
I & $0.0692(2)$ & $18(4)(6)$ & $24(10)$ & $0.30$ 
\tabularnewline
II & $0.0692(2)$ & $16(3)(3)$ & $17(10)$ & $0.64$ 
\tabularnewline
\tabularnewline
\hline\hline
$K^+ $ & $\quad a_t M \quad $ & $\quad \alpha_E^{\textrm{latt}} \quad  $   & $\quad \overline{\alpha}^{\textrm{latt}}_{EEE} \quad $ & $ 1-P $
\tabularnewline
\hline
I & $0.0971(2)$ & $8(3)(1)$ & $17(5)$ &  $0.03$
\tabularnewline
II & $0.0969(2)$ & $16(4)(3)$ & $40(9)$ &  $0.23$
\tabularnewline
\hline
\hline
\end{tabular}
\end{table}

\section{Ongoing and Future Work}
The pion polarizabilites are expected to have significant pion mass dependence, Eq.~\eqref{eq:ChPTPols}.  Additionally, because of this chiral sensitivity, the polarizabilites are also expected to be particularly sensitive to the long-range finite volume effects~\cite{Hu:2007ts}.  For these reasons, it is both interesting and crucial to perform these calculations with larger spatial volumes and lighter pion masses.  This will allow for stringent tests of the predictions of $\chi$PT as well as first-principles predictions of these staple hadronic quantities.

We are performing similar calculations for the spin-$1/2$ baryon polarizabilities~\cite{DTWL}.

\acknowledgments
This work was supported in part by the US Department of Energy.  The numerical calculations were performed with the Chroma Software Suite~\cite{Edwards:2004sx} and were carried out on the Jefferson Lab High Performance Computers.

\end{document}